\documentclass[a4paper,12pt]{article}
\pdfoutput=1

\usepackage{devanagari}

\usepackage{
graphicx,
hyperref,
amsmath,
amssymb,
charter,
xcolor,
ifluatex,
multirow}
        
\definecolor{Gray}{gray}{0.95}
\definecolor{RGray}{gray}{0.85}
\definecolor{CGray}{gray}{0.92}

\definecolor{tit}{rgb}{0.1,0.2,0.4}
\definecolor{blus}{cmyk}{1,1,0,0.6}
\definecolor{verde}{cmyk}{0.92,0,0.59,0.25}

\usepackage{tipa}
\usepackage{amsmath,amssymb,amsfonts, bm}
\usepackage{dsfont}
\usepackage{epsfig}
\usepackage{graphicx}
\usepackage{slashed}
\usepackage{color}

\usepackage{caption}
\usepackage{subcaption}
\usepackage{slashed} 

\usepackage{bbm} 
\newcommand{\eps}{\epsilon}

\newcommand{\Heff}{{\cal H}_\text{NP}}

\newcommand{\D}{{\cal D}}
\newcommand{\U}{{\cal U}}

\newcommand{\M}{{\cal M}}

\usepackage{commath}
\usepackage{calc}

\newcommand{\be}{\begin{equation}}
\newcommand{\ee}{\end{equation}}

\newcommand{\bea}{\begin{eqnarray}}
\newcommand{\eea}{\end{eqnarray}}

\newcommand{\bfig}{\begin{figure}}
\newcommand{\efig}{\end{figure}}

\newcommand{\lag}{\ensuremath{\mathcal{L}}}

\newcommand{\qqquad}{\qquad \qquad}

\makeatletter
\newcommand*{\rom}[1]{\expandafter\@slowromancap\romannumeral #1@}
\makeatother

\begin{document}
\allowdisplaybreaks
\vspace*{-2.5cm}
\begin{flushright}
{\small
IIT-BHU
}
\end{flushright}

\vspace{2cm}

\begin{center}
{\LARGE \bf \color{tit} A new solution of the fermionic mass hierarchy of
the standard model}\\[1cm]

{\large\bf Gauhar Abbas$^{a}$\footnote{email: gauhar.phy@iitbhu.ac.in}}  
\\[7mm]
{\it $^a$ } {\em Department of Physics, Indian Institute of Technology (BHU), Varanasi 221005, India}\\[3mm]

\vspace{1cm}
{\large\bf\color{blus} Abstract}
\begin{quote}
We present a new mechanism for solving the fermionic masses and mixing hierarchies
of the standard model through a minimal symmetry $\mathcal{Z}_2 \times \mathcal{Z}_5$. The mechanism is also capable of explaining the neutrino masses and mixing parameters. The phenomenological bounds arising from kaon mixing are also derived on the parameter space of the model.
\end{quote}

\thispagestyle{empty}
\end{center}

\begin{quote}
{\large\noindent\color{blus} 
}

\end{quote}

\newpage
\setcounter{footnote}{0}
\section{Introduction}
One of the most challenging puzzles of the standard model (SM) of Glashow, Salam and  Weinberg  is the observed mass pattern of the charged fermions\cite{Abbas:2017vws}.  In the SM, masses of all fermions and quark-mixing angles are arbitrary parameters, and their masses and mixing patterns are remarkably fascinating.  

For instance, the masses of the third family fermions are much larger than that of the second family, and the masses of the second family fermions are much larger than that of the first family, i.e.  $ m_\tau >> m_\mu >> m_e$, $ m_b >> m_s  >> m_d$ and $m_t  >> m_c >> m_u$.  This is the fermionic mass hierarchy among the  three fermionic families of the SM.  

There is second interesting and challenging aspect of the mass pattern of the charged fermions which is the mass hierarchy within the each family.  This mass hierarchy is really bizarre in the sense that masses of up type quarks of the second and third families are much large than that of down type quarks of same families,  on the other side, mass of the down type quark of the first family is greater than the mass of  the up type quark of the same family.  This can be written as $m_d > m_u$,   $m_c > > m_s$,  $m_t >> m_b$.  

We should also note that there is a third side of the quark mass hierarchy among the three quark families.  This is the observed mixing among the three generations of the quarks.  There is again peculiarity in the mixing pattern of the three generations of quarks in the form of the hierarchy among the quark-mixing angles,  i.e. $\sin \theta_{12} >> \sin \theta_{23} >> \sin \theta_{13}$ where $ \theta_{12}$ is the Cabibbo angle, the mixing angle between the first and second quark families, $ \theta_{23}$ is the mixing angle between the second and third quark families, and $ \theta_{13}$ is the mixing angle between the first and third quark families.

Explaining the origin of the fermionic mass hierarchy among and within the fermionic families along with the quark-mixing pattern  is a challenging problem\cite{Abbas:2017vws}-\cite{Giudice:2008uua}.  For more references, see ref.\cite{Abbas:2017vws}.

The Froggatt-Nielson mechanism is on of the most popular models for explaining the fermionic mass hierarchy and mxing of the SM\cite{Froggatt:1978nt}.  This mechanism is created by adding an  abelian flavour symmetry $U(1)_F$  to the SM in such a way that the only fermion acquiring mass through the Yukawa Lagrangian of the SM is the top quark.  The masses of  other fermions are recovered by higher dimensional operators through a flavon field charged under the  $U(1)_F$  symmetry.

This  abelian flavour symmetry $U(1)_F$ is weakly broken and is capable to distinguish  fermions among different families.  For instance, if there exists a flavon field  $\chi$ which has charge $-1$ under the abelian flavour symmetry $U(1)_F$, and charges of the fermions $\psi_i^c$ and $\psi_j$ under  $U(1)_F$ symmetry are $\theta_i$ and $\theta_j$, respectively  then  the Yukawa operator of the type $\bar{\psi}_i \varphi \psi_j$, where $\varphi$ represents the SM Higgs field, is forbidden by the  $U(1)_F$ symmetry.  However, an effective operator of the type   $\bar{\psi}_i \varphi \psi_j (\chi /\Lambda)^{(\theta_i + \theta_j)}$ is still allowed where $\Lambda$ is the scale at which new physics reveals itself.  

Thus  masses of  fermions are recovered through higher order effective operators having the following structure :
\begin{equation}
\mathcal{O} = y (\dfrac{ \chi}{\Lambda})^{(\theta_i + \theta_j)} \bar{\psi} \varphi \psi,
\end{equation}
where $y$ is the coupling constant.  The flavon field acquires a vacuum-expectation value (VEV) $\langle \chi \rangle $ which breaks the flavour $U(1)_F$ symmetry spontaneously.  

The new physics scale $\Lambda$ can be anywhere between the weak and the Planck scale.  The only essential  condition is the ratio $ \dfrac{\langle \chi  \rangle} { \Lambda}  $ should be much smaller than unity. The effect of flavon field  will be observably very small in the limit where the scale of new physics  $\Lambda$ is  larger than the weak scale.  However, the scenario where  the symmetry breaking scale is near the weak scale is promisingly interesting from the phenomenological point of view for the  high luminosity phase of the Large Hadron Collider.  Hence, the crucial question is how low this scale could be given the present bounds on flavour-changing and CP-violating processes.

This interesting question depends on the underlying unknown dynamics, for instance whether abelian flavour symmetry $U(1)_F$ is local or global.  For instance a gauged abelian flavour symmetry $U(1)_F$ can affect low energy phenomenology via exchange of the corresponding gauge boson.  If it is global and spontaneously broken the there must exists a massless Goldestone boson. 

Although a very large numbers of models have been inspired from the Froggatt-Nielson mechanism based on an abelian flavour symmetry $U(1)_F$, there are lesss  efforts to create the  Froggatt-Nielson like mechanism strictly through simple minimal abelian discrete symmetries within the minimal framework of the SM.  Such a scenario is  interesting from the theoretical point of view due to an extensive use of abelian discrete symmetries such as $\mathcal{Z}_2$ in model building, for instance two-Higgs-doublet model and minimal supersymmetric standard model.  Furthermore, as discussed earlier, low energy phenomenology is expected to take a shift since there is no local or global abelian flavour symmetry $U(1)_F$ to affect it.

Hence, in this work, we  propose a new realization of the Froggatt-Nielson mechanism within the framework of the SM  to explain origin of the observed mass pattern of fermions among and within the three fermionic families along the quark-mixing pattern  where one does not need to impose a continuous $U(1)_F$ symmetry.    Instead of a continuous abelian $U(1)_F$  symmetry, we use  two simple discrete symmetries $\mathcal{Z}_2$ and $\mathcal{Z}_5$ in the framework of the SM.

We shall proceed along the following track: In section \ref{sec2}, we discuss our new mechanism.  Neutrino masses and oscillation parameters are discussed in section \ref{sec3}. Nuemrical fits to fermion masses are presented in section \ref{sec4}.  Phenomenological bounds based on neutral kaon mixing are derived in section \ref{sec5}.  A summary of the work is presented in section \ref{sec6}. 
 \section{A new mass mechanism based on $Z_2 \times Z_{5}$} 
\label{sec2}
For achieving this  mechanism, we employ a gauge singlet flavon scalar field  $\chi$ which behaves in the following way under $SU(3)_c \times SU(2)_L \times U(1)_Y$ symmetry of the SM
\begin{eqnarray}
\chi :(1,1,0).
 \end{eqnarray} 
For this purpose two discrete symmetries  $\mathcal{Z}_2$ and $\mathcal{Z}_5$ are added  to the SM, and are imposed on the fermionic and scalar fields as shown in table\ref{tab1}. Now the only renormalized scalar-fermion coupling turns out to be  the Yukawa coupling of the top quark.  

It is noted that Yukawa couplings of other fermions are completely forbidden by the symmetries  $\mathcal{Z}_2$ and $\mathcal{Z}_5$ now.  Masses of fermions other than top quark,  are now recovered by the higher dimension operators which appears in ascending power of the expansion parameter  $ \dfrac{ \chi} { \Lambda} $. 
  \begin{table}[h]
\begin{center}
\begin{tabular}{|c|c|c|}
  \hline
  Fields             &        $\mathcal{Z}_2$                    & $\mathcal{Z}_5$        \\
  \hline
  $u_{R}, c_{R}, t_{R}$                 &   +  & $ \omega^2$                             \\
   $d_{R},  s_{R}, b_{R}, e_R, \mu_R, \tau_R$                 &   -  &     $\omega $                              \\
    $ \nu_{e_R}$                 &   -  &    $\omega^3 $                              \\
       $   \nu_{\mu_R} $                 &   -  &    $\omega^2 $                              \\
          $  \nu_{\tau_R} $                 &   +  &    $1 $                              \\
   $\psi_{L}^1$                 &   +  &    $\omega $                          \\
    $\psi_{L}^2$                 &   +  &     $\omega^4 $                         \\
     $\psi_{L}^3$                 &   +  &      $ \omega^2 $                         \\
    $\chi$                        & -  &       $ \omega$                                        \\
    $\varphi$              &   +        &     1 \\
  \hline
     \end{tabular}
\end{center}
\caption{The charges of left and  right-handed fermions of three families of the SM,right-handed neutrinos,  Higgs, and singlet scalar fields under $\mathcal{Z}_2$ and $\mathcal{Z}_5$  symmetries where $\omega$ is the fifth root of unity. }
 \label{tab1}
\end{table} 

The mass Lagrangian for fermions reads,
\bea
\label{mass1}
{\mathcal{L}}_{mass} &=& \sum_{n=0}^{2}   \left(  \dfrac{ \chi}{\Lambda} \right)^{2n}    \sum_{i,j=3,2,1}  y_{ij}^u \bar{ \psi}_{L_i}^q  \tilde{\varphi} \psi_{R_j}^{u}  + \sum_{n=0}^{2}   \left(  \dfrac{ \chi}{\Lambda} \right)^{2n+1}    \sum_{i,j=3,2,1}  y_{ij}^d \bar{ \psi}_{L_i}^q  \varphi \psi_{R_j}^{d}  \nonumber \\
&+& \sum_{n=0}^{2}   \left(  \dfrac{ \chi}{\Lambda} \right)^{2n+1}    \sum_{i,j=3,2,1}  y_{ij}^\ell \bar{ \psi}_{L_i}^\ell  \varphi \psi_{R_j}^{\ell} 
+  {\rm H.c.},
\eea
where $ \psi_{R}^u,  \psi_{R}^d, \psi_{R}^\ell    $ are right-handed up, down type singlet quarks and singlet leptons,   $ \psi_{L}^q,  \psi_{L}^\ell    $ are quark and leptonic doublets, $i$ and $j$   are family indices, $ \tilde{\varphi}= -i \sigma_2 \varphi^* $ conjugate Higgs field and $\sigma_2$ is second Pauli matrix.  We expand the Lagrangian such that it  is invariant under $\mathcal{Z}_2$ and $\mathcal{Z}_5$  symmetries.

Now  we write quark mass matrices by defining  the expansion parameter as $   \dfrac{\langle \chi \rangle} { \Lambda} = \dfrac{f}{\sqrt{2} \Lambda}= \epsilon$.  In terms of expansion parameter $\epsilon$, the up- and down-type quark mass matrices are,
\begin{equation}
\M_\U = \dfrac{v}{\sqrt{2}}
\begin{pmatrix}
y_{11}^u  \epsilon^4 &  y_{12}^u \epsilon^4  & y_{13}^u \epsilon^4    \\
y_{21}^u  \epsilon^2    & y_{22}^u \epsilon^2  &  y_{23}^u \epsilon^2    \\
y_{31}^u     &  y_{32}^u      &  y_{33}^u
\end{pmatrix}, 
\M_\D = \dfrac{v}{\sqrt{2}}
\begin{pmatrix}
y_{11}^d  \epsilon^5 &  y_{12}^d \epsilon^5 & y_{13}^d \epsilon^5   \\
y_{21}^d  \epsilon^3  & y_{22}^d \epsilon^3 &  y_{23}^d \epsilon^3  \\
 y_{31}^d \epsilon &  y_{32}^d \epsilon   &  y_{33}^d \epsilon
\end{pmatrix}.
\end{equation}
The mass matrix of charged leptons can be written as,
\begin{equation}
\M_\ell =  \dfrac{v}{\sqrt{2}}
\begin{pmatrix}
y_{11}^\ell  \epsilon^5 &  y_{12}^\ell \epsilon^5  & y_{13}^\ell \epsilon^5   \\
y_{21}^\ell  \epsilon^3  & y_{22}^\ell \epsilon^3  &  y_{23}^\ell \epsilon^3  \\
 y_{31}^\ell \epsilon   &  y_{32}^\ell \epsilon   &  y_{33}^\ell \epsilon
\end{pmatrix}.
\end{equation}
The masses of quarks and charged leptons in the approximation $\epsilon \ll 1$ can be written as\cite{Rasin:1998je},
\begin{eqnarray}
\{m_t, m_c, m_u\}  &\simeq& \{|y^u_{33}|,~ |y^u_{22}|\epsilon^2,~ |y^u_{11} -
y^u_{12}y_{21}^u/y_{22}^u|\epsilon^4\}v/\sqrt{2} ,\nonumber \\
\{m_b, m_s, m_d\} &\simeq& \{|y_{33}^d| \epsilon, ~|y_{22}^d| \epsilon^3,~
|y_{11}^d-y_{12}^d y_{21}^d/y_{22}^d| \epsilon^5\}v/\sqrt{2}, 
\nonumber \\
\{m_\tau, m_\mu, m_e\} &\simeq& \{|y_{33}^\ell| \epsilon, ~|y_{22}^\ell| \epsilon^3,~
|y_{11}^\ell|\epsilon^5\}v/\sqrt{2}.
\end {eqnarray}
Similarly the quark mixing angles at leading order are found to be\cite{Rasin:1998je},
\begin{eqnarray}
\sin \theta_{12}  \simeq |V_{us}| &\simeq& \left|{y_{21}^d \over y_{22}^d}  -{y_{21}^u \over y_{22}^u}  \right| \epsilon^2, 
\sin \theta_{23}  \simeq |V_{cb}| \simeq  \left|{y_{32}^d \over y_{33}^d}  -{y_{32}^u \over y_{33}^u}  \right|  \epsilon^2,\nonumber \\
\sin \theta_{13}  \simeq |V_{ub}| &\simeq& \left|{y_{31}^d \over y_{33}^d}  -{y_{21}^u y_{32}^d \over y_{22}^u y_{33}^d} 
- {y_{31}^u \over y_{33}^u} \right|  \epsilon^4.
\end{eqnarray}
From the above results, we observe that  $ \sin \theta_{13} $ is much suppressed relative to $ \sin \theta_{23} $ and $ \sin \theta_{12} $.  Thus, this model explains the quark-mixing pattern in a satisfactory way. We note that the $\sin \theta_{23} $ is of same order as $\sin \theta_{23}$.   Similar results for the quark mixing angles are reported in ref.\cite{Babu:1999me}.
\section{Neutrino masses and oscillations}
\label{sec3}
For recovering neutrino masses and oscillation parameters, we add three right-handed neutrinos to the model as  shown in table \ref{tab1}.  The tree level Majorana Lagrangian can be written by observing the charges of the right-handed neutrinos under $\mathcal{Z}_2$ and $\mathcal{Z}_5$  symmetries given in table \ref{tab1}, 
\bea
\mathcal{L}_M  = M_1 \bar{\nu^c}_{e_R} \nu_{\mu_R} + M_2 \bar{\nu^c}_{\mu_R} \nu_{e_R} + M_3 \bar{\nu^c}_{\tau_R} \nu_{\tau_R},
\eea
where $M_{1,2,3}$ are the Majorana mass scales.

The non-renormalizable Majorana Lagrangian is written using the charges of the flavon field and  the right-handed neutrinos under $\mathcal{Z}_2$ and $\mathcal{Z}_5$  symmetries given in  table \ref{tab1}. Thus the Lagrangian is,
\bea
\label{mass1}
{\mathcal{L}}_{mass} &=& 
    \left(  \dfrac{ \chi}{\Lambda} \right)^{4}    c_{11}^\nu  \bar{\nu^c}_{e_R}  \nu^c_{e_R}
+   \left(  \dfrac{ \chi^\dagger}{\Lambda} \right)^{4}    c_{22}^\nu  \bar{\nu^c}_{\mu_R}  \nu^c_{\mu_R} +   \left(  \dfrac{ \chi^\dagger}{\Lambda} \right)^{3}    c_{13}^\nu  \bar{\nu^c}_{e_R}  \nu^c_{\tau_R}  \\ \nonumber 
&+&  \left(  \dfrac{ \chi^\dagger}{\Lambda} \right)^{3}    c_{23}^\nu  \bar{\nu^c}_{\mu_R}  \nu^c_{\tau_R} +  \left(  \dfrac{ \chi^\dagger}{\Lambda} \right)^{3}    c_{13}^\nu  \bar{\nu^c}_{\tau_R}  \nu^c_{e_R} +  \left(  \dfrac{ \chi^\dagger}{\Lambda} \right)^{3}    c_{23}^\nu  \bar{\nu^c}_{\tau_R}  \nu^c_{\mu_R}.
\eea
The neutrino Dirac mass matrix is given by, 
\begin{equation}
\label{NM}
\M_{\D} = \dfrac{v}{\sqrt{2}}
\begin{pmatrix}
y_{11}^\nu  \epsilon^3 &  y_{12}^\nu \epsilon & y_{13}^\nu   \epsilon^{4}  \\
y_{21}^\nu  \epsilon  & y_{22}^\nu \epsilon^3 &  y_{23}^\nu   \epsilon^4 \\
y_{31}^\nu \epsilon   &  y_{32}^\nu  \epsilon^5   &  y_{33}^\nu \epsilon^2
\end{pmatrix}.
\end{equation}
The neutrino mass matrix after including the Majorana mass terms becomes,
\begin{equation}
\label{NM}
\M = 
\begin{pmatrix}
0 &  \M_{\D} \\
\M_{\D}^T & \M_{R}   \\
\end{pmatrix},
\end{equation}
where the Majorana mass matrix $\M_R$ is,
\begin{equation}
\label{NM}
\M_{R} = 
\begin{pmatrix}
c_{11}^\nu  \epsilon^4 &  M & c_{13}^\nu  \epsilon^3 \\
  M  & c_{22}^\nu \epsilon^4 & c_{23}^\nu \epsilon^3 \\
c_{13}^\nu  \epsilon^3   & c_{23}^\nu \epsilon^3   & M
\end{pmatrix},
\end{equation}
where $M_1=M_2=M_3 = M$ is assumed for the simplification.

The masses of neutrinos  now can be determined using type-\rom{1} seesaw mechanism\cite{seesaw1}-\cite{seesaw5}. Assuming $ \M_{\D} << \M_{R}$, the mass matrix of the light neutrinos reads,
\begin{eqnarray}
{ \M}~ =~ -  \M_{\D}  \M_{R}^{-1}  \M_{\D}^T.
\end{eqnarray}
The light neutrino masses can  approximately be written as\cite{Rasin:1998je},
\bea
m_1 &\approx &   y_{11}^\nu \epsilon^2 \epsilon^\prime, ~ m_2 \approx    y_{22}^\nu \epsilon~ \epsilon^\prime,~ m_3 \approx y_{33}^\nu \epsilon ~ \epsilon^\prime,
\eea 
where $\epsilon^\prime = \frac{v}{\sqrt{2} M} $.

For determining the leptonic mixing angles,  the following unitary transformation  is used to diagonalize the charged lepton mass matrix: 
\begin{eqnarray}
(V^\ell_L)^\dagger \M_\ell (\M_\ell)^\dagger V^\ell_L &=&{\rm diag}(m_e^2,m_\mu^2,m_\tau^2).
\end{eqnarray}
The leptonic mixing is parmetrized by the Pontecorvo-Maki-Nakagawa-Sakata (PMNS) matrix defined as   $U_{\rm PMNS}=(V^\ell_L)^\dagger V^\nu_L$,
where $V^\nu_L$ is used to diagonalize the  neutrino mass matrix via,
\begin{eqnarray}
(V^\nu_L)^\dagger \M(\M)^\dagger V^\nu_L &=& {\rm diag}(m_1^2,m_2^2,m_3^2).
\end{eqnarray}

The leading form of $V^\ell_L$  is found to be
\begin{equation}
V^\ell_L=\left(\begin{array}{ccc}
1 & \frac{y^\ell_{12}}{y^\ell_{22}}\epsilon^2 & \frac{y^\ell_{13}}{h^\ell_{33}}\epsilon^4 \\
-\frac{y^\ell_{12}}{y^\ell_{22}}\epsilon^2 & 1 & \frac{y^\ell_{23}}{y^\ell_{33}}\epsilon^2 \\
-\frac{y^\ell_{13}y^\ell_{22}-y^\ell_{23}y^\ell_{12}}{y^\ell_{22}y^\ell_{33}}\epsilon^4 &
-\frac{y^\ell_{23}}{y^\ell_{33}}\epsilon^2 & 1
\end{array}\right).
\end{equation}
We observe that there is at least $\epsilon^2$ order suppression of the  off-diagonal elements in the above matrix.  Therefore,  the matrix $V^\ell_L$  can be approximately taken close to a unit matrix and, the PMNS matrix is practically dominated by the unitary matrix $V^\nu_L$.

The leptonic mixing angles approximately can be read as\cite{Rasin:1998je},
\begin{eqnarray}
\sin \theta_{12}  &\simeq& \left|{y_{21}^\nu \over y_{22}^\nu}  \right| \epsilon^2, 
\sin \theta_{23} \simeq  \left|{y_{32}^\nu \over y_{33}^\nu}  \right|  ,\nonumber 
\sin \theta_{13} \simeq \left|{y_{31}^\nu \over y_{33}^\nu} \right|  \epsilon^2.
\end{eqnarray}
It is interesting to note that $\sin \theta_{13}$ is naturally small and is close to the Cabibbo angle as observed in experiments.  On the other side, $\sin \theta_{23}$ is unsuppressed as required by experimental findings.
\section{Fitting fermionic masses, quark-mixing and neutrino oscillation parameters}
\label{sec4}
We reproduce the fermion masses using the following values of the fermion masses at $\mu= f= 1$TeV\cite{Xing:2007fb},
\begin{eqnarray}
\{m_t, m_c, m_u\} &\simeq& \{150.7 \pm 3.4,~ 0.532^{+0.074}_{-0.073},~ (1.10^{+0.43}_{-0.37}) \times 10^{-3}\}~{\rm GeV}, \nonumber \\
\{m_b, m_s, m_d\} &\simeq& \{2.43\pm 0.08,~ 4.7^{+1.4}_{-1.3} \times 10^{-2},~ 2.50^{+1.08}_{-1.03} \times 10^{-3}\}~{\rm GeV},
\nonumber \\
\{m_\tau, m_\mu, m_e\} &\simeq& \{1.78\pm 0.2,~ 0.105^{+9.4 \times 10^{-9}}_{-9.3 \times 10^{-9}},~ 4.96\pm 0.00000043 \times 10^{-4}\}~{\rm GeV}.
\end{eqnarray}
The magnitudes and phases  of the CKM mixing elements are \cite{Tanabashi:2018oca},
\bea
|V_{us}| &=& 0.97446 \pm 0.0004,  |V_{cb}| = 0.04214 \pm 0.00075, |V_{ub}| = 0.00365 \pm 0.00012, \\ \nonumber
\sin 2 \beta &=& 0.691 \pm 0.017, ~ \alpha = (84.5^{+5.9}_{-5.2})^\circ,~  \gamma = (73.5^{+4.2}_{-5.1})^\circ.
\eea
The present scenario of the neutrino physics for the normal hierarchy can be described by the following global fit results\cite{deSalas:2017kay},
\bea
\Delta m_{21}^2 &=& (7.55^{+0.59}_{-0.5}) \times 10^{-5} {\rm eV}^2, |\Delta m_{31}^2| = (2.50\pm 0.09) \times 10^{-3} \rm{eV}^2,  \\ \nonumber
\sin^2 \theta_{12} &=&  (3.20^{+0.59}_{-0.47}) \times 10^{-1},
 \sin^2 \theta_{23} =  (5.47^{+0.52}_{-1.02}) \times 10^{-1},  \sin^2 \theta_{13} =  (2.160^{+0.25}_{-0.20}) \times 10^{-2},
\eea
where range of errors is $3 \sigma$.

The purpose of the fit is to show that the model can reproduce the fermion masses in a realistic way. We choose to fit the mass expressions of fermions  provided by the model to the fermion masses at TEV scale  keeping in mind related phenomenological investigation. In the phenomenological investigation, the flavon contribution to K-K oscillation is at tree level, and top quark and other fermions do not contribute here.  Therefore, we do not run them to the  eletroweak scale.  However, in some other processes where top quark is involved, this running may be essential.

We note that the masses of fermions are derived using bi-unitary transformation which allows the masses to be positive.  The discovered Higgs boson at the LHC is placing tighter constraints on the sign of the down type quarks.  However, both signs are still alowed.  We have scanned the parameters space of our model in the most general way keeping both signs for Yukawa couplings.

We first fit  quark masses by defining $\chi^2 = \dfrac{(m_q - m_q^{\rm{model}} )^2}{\sigma_{m_q}^2}+  \dfrac{(m_\ell - m_\ell^{\rm{model}} )^2}{\sigma_{m_\ell}^2}  + \dfrac{(\sin \theta_{ij} -\sin \theta_{ij}^{\rm{model}} )^2}{\sigma_{\sin \theta_{ij}}^2}     + \dfrac{(\sin 2 \beta  -\sin 2 \beta^{\rm{model}} )^2}{\sigma_{\sin2\beta}^2}   + \dfrac{( \alpha  - \alpha^{\rm{model}} )^2}{(\sigma_{\alpha})^2}  + \dfrac{( \gamma   - \gamma^{\rm{model}} )^2}{(\sigma_{\gamma})^2}$ where $q=u,d,c,s,t,b$, $\ell=e,\mu,\tau$ and $i,j=1,2,3$.  The phases of the CKM matrix in the standard choice are defined as followos:
\begin{eqnarray}
\beta^{\text{model}} =\text{arg} \left(- \dfrac{V_{cd} V_{cb}^*}{V_{td} V_{tb}^*}\right),~\alpha^{\text{model}} =\text{arg} \left(- \dfrac{V_{td} V_{tb}^*}{V_{ud} V_{ub}^*}\right),~\gamma^{\text{model}} =\text{arg} \left(- \dfrac{V_{ud} V_{ub}^*}{V_{cd} V_{cb}^*}\right).
\end{eqnarray}
The dimensionless coefficients $y_{ij}^{u,d,\ell}= |y_{ij}^{u,d,\ell}| e^{i \phi_{ij}^{q,\ell}}$  are scanned with $|y_{ij}^{u,d,\ell}| \in [0.1,2\pi]$ and $ \phi_{ij}^{q,\ell} \in [0,2\pi]$.  The fit results are,
\begin{eqnarray}
\{|y_{33}^u|, |y_{22}^u|, |y_{11}^u-y_{12}^u y_{21}^u/y_{22}^u|\} &\simeq& \{0.87, 0.3,
0.1\}, \nonumber \\
\{|y_{33}^d|, |y_{22}^d|, |y_{11}^d-y_{12}^d y_{21}^d/y_{22}^d|\} &\simeq& \{0.14, 0.29, 1.44\}, \nonumber \\
\{|y_{33}^\ell|, |y_{22}^\ell|, |y_{11}^\ell|\} &\simeq& \{0.1, 0.6, 0.29\}, \nonumber \\
\{|y_{32}^u|, |y_{31}^u|, |y_{21}^u|,|y_{12}^u| \} &\simeq& \{2.29, 1.71,
1.46, 0.26\}, \nonumber \\
\{\phi_{33}^u, \phi_{32}^u,  \phi_{31}^u,\phi_{22}^u,\phi_{21}^u,\phi_{12}^u,\phi_{11}^u \} &\simeq& \{ 
3.34,  3.96, 2.99, 1.0, 5.11, 4.37, 2.19\}, \nonumber \\
\{|y_{32}^d|, |y_{31}^d|,  |y_{21}^d|,|y_{12}^d| \} &\simeq& \{0.36, 3.67, 5.16, 0.13\}, \nonumber \\
\{\phi_{33}^d, \phi_{32}^d,  \phi_{31}^d,  \phi_{22}^d,\phi_{21}^d,\phi_{12}^d,\phi_{11}^d \} &\simeq& \{ 
3.05, 1.79, 3.0, 2.63, 2.71, 3.15, 3.82     \}, \nonumber \\
\epsilon & = & 0.1, \delta \approx 1.2, ~\chi^2_{min} \approx 4.4,
\end{eqnarray}
where $\delta$ is Dirac $CP$ phase.

Now we define  
$\chi^2 =   \dfrac{(\Delta m_{21}^2 - \Delta m_{21}^{2 ~ \rm{model}} )^2 }{\sigma_{\Delta m_{21}^2}^2} + \dfrac{(\Delta m_{31}^2 - \Delta m_{31}^{2 ~ \rm{model}} )^2 }{\sigma_{\Delta m_{31}^2}^2}  + \dfrac{(\sin \theta_{ij}^\nu -\sin \theta_{ij}^{\nu ~\rm{model}} )^2}{\sigma_{\sin \theta_{ij}^\nu}^2}  $ where $i,j=1,2,3$ for fitting the neutrino oscillation data.     The result of fitting is\footnote{The Majorana scale for this fit can vary between $10^7$ to $10^{11}$ GeV.},
\begin{eqnarray}
\{y_{33}^\nu, y_{22}^\nu, y_{11}^\nu\, y_{32}^\nu, y_{31}^\nu, y_{21}^\nu\} & = & \{0.41, 0.10,
0.75, 0.3,5.94,  5.74 \}, \nonumber \\
\epsilon^\prime & = & 1.259 \times 10^{-9}, \chi^2_{min}  =  1.79.
\end{eqnarray}
\section{Phenomenological analysis}
\label{sec5}
The scalar potential of the model acquires the form,
\begin{align}
- \lag_\text{potential}
= - \mu_\chi^2\, \chi^*  \chi
  + \lambda_\chi\, (\chi^* \chi)^2 
  + \rho \, (\chi^2 + \chi^{* 2} ) 
  + \lambda_{\varphi  \chi}  (\chi^* \chi)  (\varphi^\dagger \varphi)+V(\varphi),
\label{eq:potential}
\end{align}
where the third term causes the soft-breaking of the $\mathcal{Z}_5$ symmetry.  We work in the limit where $ \lambda_{\varphi  \chi}   =0$\cite{Bauer:2016rxs}\footnote{The condition $\lambda_{\phi \chi} = 0$ is adopted purely from a phenomenlogical point since the Higgs discovered by the LHC is behaving like a SM Higgs.  The contribution of this coupling to loop-level FCNC process will be suppressed by loop factor relative to the tree level  flavon contribution to the FCNC process in the absence of this coupling. Hence, it can be ignored.}.   The non-zero values of  $ \lambda_{\varphi  \chi} $ will give rise to Higgs-flavon mixing\cite{Berger:2014gga}.  

After symmetry breaking, the  flavon field is given  by excitations around its VEV,
\begin{align}
 \chi(x)=\frac{f + s(x) +i\, a(x)}{\sqrt{2}}.
\label{eq:flavonfield}
\end{align}
The scalar and pseudo-scalar  components have the following masses:
\begin{align}
m_s = \tilde{\mu}_\chi\,= \sqrt{\lambda_\chi} f 
\qqquad \text{and} \qqquad
m_a= \sqrt{2 \rho},
\label{eq:masses}
\end{align}
where  $ \tilde{\mu}_\chi= \sqrt{\mu_\chi- 2 \rho}$.  Assuming the mass of the pseudo-scalar below the flavor scale,  we can assume the mass hierarchy
\begin{align}
m_a < m_s \approx f < \Lambda.
\end{align}
This is done purely for the phenomenological purpose and means that psuedoscalae flavon is lighter than the scalar one and the scale which renormalizes the model.\footnote{One of the reason to do so is to compare our results with the reference \cite{Bauer:2016rxs} which also uses same assumption.}

The couplings of the flavon field with fermions can be written as\footnote{This can be done by using the parametrization of the flavon field after symmetry breaking, for example, when we expand 12 term in down type quark matrix which is $y_{12}^d (f + s + i a)^5/\sqrt{2}$, we get the coupling of pseudoscalar flavon to d and s quarks which is $5 y_{12}^d \epsilon^5$.  
},
\begin{eqnarray}
\label{fup}
y_{af_{iL} f_{jR}}^{u} &\equiv & y_{aij}^u = \frac{1}{f} 
\begin{pmatrix}
4 y_{11}^u  \epsilon^4 & 4 y_{12}^u \epsilon^4  &  4 y_{13}^u \epsilon^4    \\
2 y_{21}^u  \epsilon^2    & 2 y_{22}^u \epsilon^2  &  2 y_{23}^u \epsilon^2     \\
0   & 0   & 0
\end{pmatrix},
y_{aij}^{d} = \frac{1}{f} 
\begin{pmatrix}
5 y_{11}^d  \epsilon^5 & 5 y_{12}^d \epsilon^5 & 5 y_{13}^d \epsilon^5   \\
3 y_{21}^d  \epsilon^3  & 3 y_{22}^d \epsilon^3 & 3 y_{23}^d \epsilon^3  \\
  y_{31}^d \epsilon &   y_{32}^d \epsilon   &  y_{33}^d \epsilon
\end{pmatrix}.
\end{eqnarray}
For the pseudoscalar component of flavon field, the following notation is used:
\begin{equation}
y_{ij}= y_{sf_{iL} f_{iR}} = i  y_{af_{iL} f_{iR}}. 
\end{equation}
Neutral meson mixing places one of the most stringent constraints on flavon couplings to fermions. Flavon couplings to fermions are sources of flavor-changing neutral currents which are tightly constrained by meson anti-meson mixing. The effective Hamiltonian describing $\Delta F =2$ interactions can be written as,
\begin{align}
\Heff^{\Delta F=2}&=C_1^{ij} \,( \bar q^i_L\,\gamma_\mu \, q^j_L)^2+\widetilde C_1^{ij} \,( \bar q^i_R\,\gamma_\mu \, 
q^j_R)^2 +C_2^{ij} \,( \bar q^i_R \, q^j_L)^2+\widetilde C_2^{ij} \,( \bar q^i_L \, q^j_R)^2\notag\\
&+ C_4^{ij}\, ( \bar q^i_R \, q^j_L)\, ( \bar q^i_L \, q^j_R)\,+C_5^{ij}\, ( \bar q^i_L \,\gamma_\mu\, q^j_L)\, ( \bar q^i_R \,
\gamma^\mu q^j_R)\,+ \text{h.c.}.
\label{eq:heffdf2}
\end{align}
The tree-level  contribution of flavon contributes to the following Wilson coefficients is\cite{Buras:2013rqa,Crivellin:2013wna},
\begin{align}
C_2^{ij} &= -(y_{ji}^*)^2\left(\frac{1}{m_s^2}-\frac{1}{m_a^2}\right),
\tilde C_2^{ij} = -y_{ij}^2\left(\frac{1}{m_s^2}-\frac{1}{m_a^2}\right),
C_4^{ij} = -\frac{y_{ij}y_{ji}}{2}\left(\frac{1}{m_s^2}+\frac{1}{m_a^2}\right)\,.
\label{eq:wilsons}
\end{align}
\begin{figure}
	\centering
		\includegraphics[width=6.0cm, height=6cm]{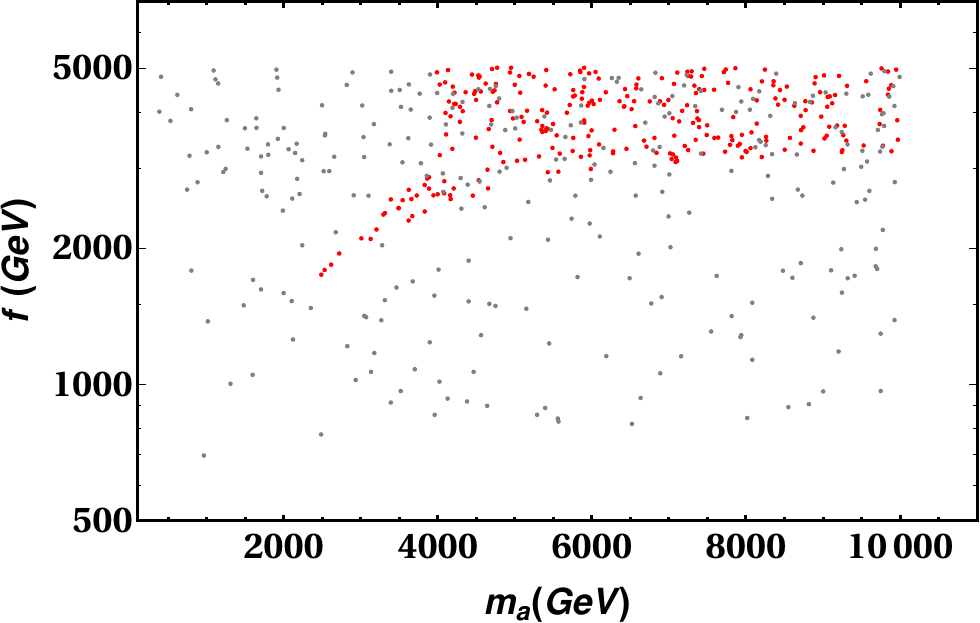}
\caption{The allowed parameter space by $\epsilon_K$ and $\Delta m_K$ for $\lambda_\chi= 2$ in the $m_a - f$ plane.  The red points represent allowed flavon contribution to $\epsilon_K$, and the allowed contribution to $\Delta m_K$ is shown by grey points.  }\label{fals2}
\end{figure}
The RG running of the Wilson coefficients and the matrix elements are adopted from Refs.~\cite{Bona:2007vi,Ciuchini:1998ix}.  

In this work, we present bounds on the parameter space from the neutral kaon mixing.  The theoretical errors dominate in the $CP$ violating parameter $\epsilon_{K}$ of kaon mixing because of its remarkable experimental precision.  In the case of $B$ meson mixing, flavon coupling, as can be seen from eq.\ref{fup}, is less suppressed in comparison to kaon mixing.  The $D$ meson mixing is affected by the hadronic uncertainties.  Moreover, it is less suppressed relative to kaon mixing as can be seen from the eq. \ref{fup}. 

The following parameters  with 95\%~CL limits from $K-\bar K$ mixing are used in deriving our bounds~\cite{Bona:2007vi}:
\begin{align}
C_{\eps_K}&=\frac{ \text{Im} \langle K^0|\mathcal{H}^{\Delta F=2}|\bar K^0\rangle}{\text{Im} \langle K^0| \mathcal{H}_\text{SM}^{\Delta F=2} |\bar K^0 \rangle} = 1.05_{-0.28}^{+0.36},
C_{\Delta m_K} =\frac{\text{Re}\langle K^0|\mathcal{H}^{\Delta F=2}|\bar K^0\rangle}{\text{Re} \langle K^0| \mathcal{H}_\text{SM}^{\Delta F=2} |\bar K^0 \rangle} = 0.93_{-0.42}^{+1.14} ,
\end{align}
where $\mathcal{H}^{\Delta F=2}$ parametrizes the SM and flavon
contributions, and  $\mathcal{H}_\text{SM}^{\Delta F=2}$ stands only for the SM contribution. The SM contribution is directly taken from Ref. \cite{Crivellin:2013wna}.

In the fig.\ref{fals2}, we show bounds in the $m_a - f$ plane derived by constraints coming from  $\epsilon_K$ and $\Delta m_K$ where we have chosen $\lambda_\chi= 2$.  As it is obvious, $\epsilon_K$ places tighter constraints on the parameter space which are represented by the red points.

\section{Summary}
\label{sec6}
We have discussed a  new Froggatt-Nielsen mechanism as a solution of the fermionic hierarchy of the SM which is based on the discrete symmetries  $\mathcal{Z}_2$ and $\mathcal{Z}_5$.  The  main feature of the model, besides its simplicity which is exhibited by the mass matrices of fermions, is that a very minimal symmetry  $\mathcal{Z}_2 \times \mathcal{Z}_5 $ (probably the smallest known symmetry)  can privide an explaination for the whole fermionic spectrum of the SM including neutrino masses and oscillations.

A partial phenomenological analysis is also presented which will be extended further in a future work. We have also shown that neutrino masses and oscillations parameters can also be recovered.

On the theoretical side, this mechanism is different from the standard Froggatt-Nielsen mechanism in the sense that it is based on the discrete symmetries $\mathcal{Z}_2$ and $\mathcal{Z}_5$ although it still employs a singlet gauge scalar field like the conventional mechansim.

We note that  an origin of  $\mathcal{Z}_2$ and  $\mathcal{Z}_5$  may be traced to Abelian or non-Abelian continous symmetries.  For instance, $\mathcal{Z}_2$ and $\mathcal{Z}_5$   may be an artefact of spontaneous  breaking of $U(1) \times  U(1)$  continuous symmetries.  
\section*{Acknowledgement}
I am extremely grateful to Prof. Anjan S. Joshipura for important comments and suggestions on this work.

\end{document}